\documentclass[a4paper,3p]{elsarticle}
\journal{Journal of Computational Physics}
\usepackage{hyperref}
\usepackage{graphicx,caption}
\usepackage{cmap}
\usepackage[T2A]{fontenc}
\usepackage[utf8]{inputenc}
\usepackage[english]{babel}
\usepackage{textcomp}
\usepackage{xcolor}
\usepackage{amsmath}
\usepackage{amssymb}
\usepackage{latexsym}
\usepackage[normalem]{ulem}
\usepackage{physics}
\usepackage{multirow}
\usepackage{hhline}
\usepackage{subfigure}
\usepackage{comment}
\usepackage{float}
\usepackage{multicol}
\usepackage{blindtext}
\begin{document}

\begin{frontmatter}

\title{Adjusting the numerical viscosity in the Godunov-like SPH method at modeling compressible flows}

\author{A.~N.~Parshikov}
\ead{anatoliy.parshikov@gmail.com}
\author{S.~A.~Medin}
\author{G.~D.~Rublev}
\ead{rublev_gd_97@vk.com}
\author{S.~A.~Dyachkov}
\ead{sergei.dyachkov@phystech.edu}

\address{Dukhov Research Institute of Automatics, Moscow 127030, Russia}
\address{Joint Institute for High Temperatures of RAS, Moscow, 125412, Russia}

\begin{abstract}
The paper proposes a way to control the viscosity of numerical approximation in the contact SPH method. This variant of SPH contains momentum and energy fluxes in the right-hand sides of the equations, which are calculated using the solution of the Riemann problem between each pair of neighboring particles within the smoothing kernel support radius, which is similar to the procedure for calculating fluxes across cell boundaries in Godunov schemes. Such SPH method does not require the use of artificial viscosity, because the significant numerical viscosity is already introduced by a Riemann problem solution. The magnitude of numerical viscosity decreases linearly with particle size, however, it becomes comparable with the physical viscosity for most materials when the particle size is about $\thicksim 1\,$nm, which hampers the correct accounting for viscous effects in real-life problems. In this study we develop a method for reducing the viscous stresses of numerical origin, for which a correcting viscous stress tensor is constructed on the basis of the analytical solution for discontinuous viscous flow. The use of such correction makes it possible to improve the agreement with the experiment in the simulation of viscous flows.
\end{abstract}


\end{frontmatter}

\section{Introduction}

In recent years the range of applications for meshless smoothed particle hydrodynamics (SPH) method has grown dramatically: being developed for astrophysical problems~\cite{Monaghan:MNRAS:1977} it is now an essential tool in variety of engineering packages along with the well-established Eulerian or Lagrangian mesh-based schemes. Due to its meshless Lagrangian nature, SPH is successfully applied to model flows with complex free boundaries~\cite{Monaghan:JCP:1994}, inhomogeneous multi-material flows~\cite{Colagrossi:JCP:2003,DualSPHysics2021}, or problems with extreme strains and fracture of solids~\cite{Dyachkov:JAP:2019}. It is worth noting that implementation and parallelization of SPH is less complicated than of mesh-based methods, and the variety of efficient parallel algorithms has been developed~\cite{DualSPHysics2021,Egorova:CPC:2019,Zhu:CPC:2023}.

Simulations of viscous flows are demanded in various applications, and SPH is no exception. For our knowledge, the viscous stress tensor was first introduced to classical SPH scheme by Monaghan and Gingold~\cite{Monaghan:JCP:1983}, but with the purpose of stabilizing the propagation of shock waves with the artificial viscosity. Morris and Monaghan~\cite{Morris:JCP:1997:1} has also developed an approach, which adaptively changes the artificial viscosity, so that it increases at shocks and reduces at other regions. Takeda et al.~\cite{Takeda:PTP:1994} introduced viscosity tensor to weakly-compressible SPH for solving the real Navier--Stokes equations: they successfully modeled Poiseuille flow in a cylinder, flow around a cylinder, and developed the corresponding boundary conditions. Morris et al.~\cite{Morris:JCP:1997:2} introduced the viscosity tensor to incompressible SPH and tested it for similar problems at low Reynolds numbers, while Lastiwka et al.~\cite{Lastiwka:IntJNumMethodsFluids:2009} developed an improved inflow/outflow boundary condition for the problem with the flow around a cylinder. Marrone et al.~\cite{Marrone:JCP:2013} proposed a scheme to model viscous SPH flows for Reynolds numbers ranging from 10 to 2400, demonstrating the formation of vortices in the flows passing an obstacle.

In the above studies the classical or weakly-compressible SPH formulation was used. The Godunov-like (contact) SPH method is proposed for modeling nonsteady flows of viscous compressible fluids: in the right-hand sides of SPH equations the momentum and energy fluxes are calculated using the Riemann problem solution between the each pair of neighbor particles within the smoothing kernel width. This method was first published in Ref.~\cite{Parshikov:CMMP:1999}, where the equations of the contact SPH method are given with an interparticle Riemann problem solution for the compressible inviscid gas or fluid. The generalization of the method for elastic-plastic media and two-dimensional axis-symmetric geometry was provided in Ref.~\cite{Parshikov:IJIE:2000}. The generalization of the method to 3D flows of heat-conductive elastic-plastic media is done in~\cite{Parshikov:JCP:2002}, where a complete set of equations is given with the relations for the thermal contact between SPH particles.

The idea of evaluating the heat diffusion at contacts with discontinuities proposed in Ref.~\cite{Cleary:JCP:1999} has also been developed in Ref.~\cite{Parshikov:JCP:2002}. Similar to the Riemann problem for hydrodynamic set of equations, the analytical solution of temperature discontinuity breakup was introduced to provide a better contact temperature approximation. Viscosity, similar to heat conduction, results in the momentum diffusion which can be expressed using similar approach, which is introduced in this study. Thus, we complete the study~\cite{Parshikov:JCP:2002} by formulating the SPH contact method for calculating flows of compressible viscoelastic-plastic media with thermal conductivity.

The main problem of the proposed SPH method is large numerical viscosity induced by a Riemann problem solution, which complicates simulation of fluids with realistic viscosity. It should be noted, that the problem of numerical viscosity origin in various fluid dynamics methods is studied for decades~\cite{Yanilkin:2016}. In contrast to Monaghan and Gingold~\cite{Monaghan:JCP:1983}, who introduced the viscosity tensor to stabilize the solution of SPH equations, in our formulation the viscosity tensor may be used to reduce the numerical viscosity of the Godunov-like SPH methods. Such approach allows to model flows with realistic viscosity which can be obtained by either increase or decrease the numerical viscosity. However, the latter requires the evaluation of the numerical viscosity magnitude, so that we provide the technique for numerical viscosity measurement based on shear test simulation and construct the analytical approximation for it. To be sure of the validity of our numerical viscosity control procedure, we simulate the Stokes experiment with falling ball and compare the calculated value of the numerical viscosity of the fluid with its value determined from the shear test.

\section{The origin of numerical viscosity in the Godunov-like SPH}

The flow of an inviscid compressible fluid is described by the set of equations expressing the conservation laws for the mass, the momentum, and the energy:
\begin{equation}
\label{Euler1}
\frac{1}{\rho}\frac{d\rho}{dt} = - \dot{\varepsilon}= - \nabla \cdot \overrightarrow{U},
\end{equation}
\begin{equation}
\label{Euler2}
\frac{d\overrightarrow{U}}{dt} = -\frac{1}{\rho}\nabla P,
\end{equation}
\begin{equation}
\label{Euler3}
\frac{dE}{dt} = -\frac{1}{\rho}\nabla \cdot (P \overrightarrow{U}),
\end{equation}
where $\overrightarrow{U}$ is velocity, $E=e+\overrightarrow{U}^2/2$ is sum of internal and kinetic energy per unit mass, i.e. total energy per unit mass, $\dot{\varepsilon}$ is volumetric strain rate, $P$ is pressure, $\rho$ is density. The system \eqref{Euler1}--\eqref{Euler3} is closed with an equation of state. Simulations presented in our study are performed with the linear equation of state, which connects the pressure $P$ with the density of matter $\rho$:
\begin{equation}
\label{Grun}
P= C^2(\rho-\rho_0),
\end{equation}
where $C$ is speed of sound.

\begin{figure}[t]
	\centering
	\includegraphics[width=1.0\textwidth]{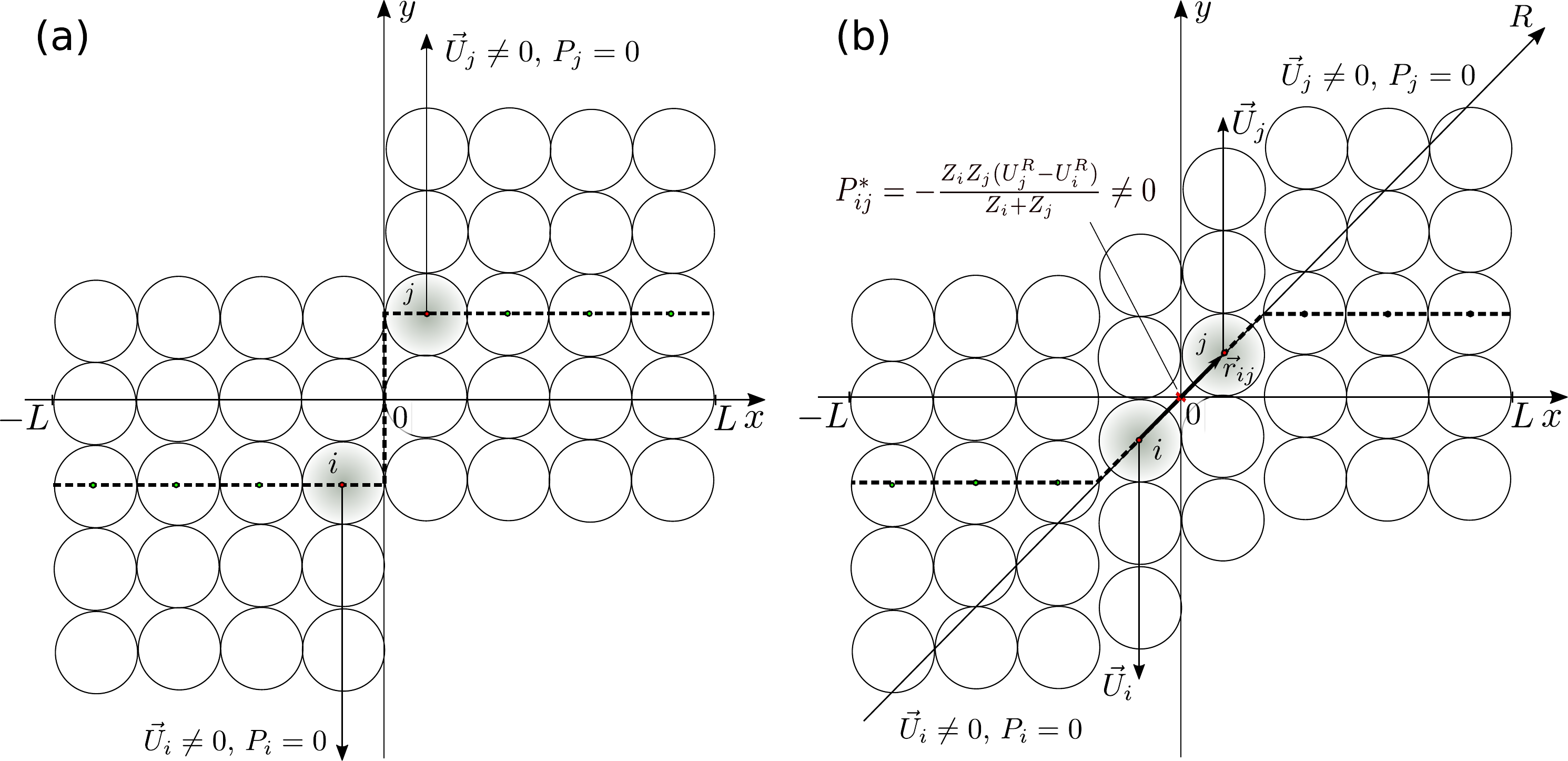}
	\caption {\label{sph-viscosity-scheme}
		The scheme of mutual displacement of inviscid flows in opposite directions with the interface along the plane $x=0$: (a) the momentum diffusion is absent; (b)
		the momentum diffusion appears as a result of SPH simulation with the contact pressure approximation. The velocity dependence in Eq.~\eqref{RazpadP} at $P_i = P_j = 0$ results in momentum transfer. The dotted lines show the evolution of the control surface passing through the centers of particles, which initially (at $t=0$) coincide with the $x$-axis.}
\end{figure}

The SPH approximation of the equations (\ref{Euler1})--(\ref{Euler3}) by the SPH contact method \cite{Parshikov:JCP:2002} has the following form:
\begin{equation}
\label{SPHmass1}
\frac{d\varepsilon_i}{dt}=-2\sum\limits_j \frac{m_j}{\rho_j}\left(U_i^R - U_{ij}^{*R}\right)\overrightarrow{e^R}\cdot \nabla_iW_{ij},
\end{equation}
\begin{equation}
\label{SPHmomentum1}
\frac{d\overrightarrow{U}_i}{dt}=-2\sum\limits_j \frac{m_j P^{*}_{ij}}{\rho_j\rho_i}\nabla_iW_{ij},
\end{equation}
\begin{equation}
\label{SPHenergy1}
\frac{dE_i}{dt}=-2\sum\limits_j \frac{m_jP^{*}_{ij}U_{ij}^{*R}}{\rho_j\rho_i}\overrightarrow{e^R}\cdot \nabla_iW_{ij},
\end{equation}
where the smoothing kernel $W_{ij}$ is a function of $\left|\vec{r}_{ij}\right|/h$ ($h$ is the smoothing distance), $P^{*}_{ij}$ and $U_{ij}^{*R}$ are the pressure and velocity obtained as a solution of the Riemann problem at the contact between particles $i$~and~$j$. In the acoustic approximation, these quantities are
\begin{equation}
\label{RazpadP}
P^{*}_{ij} = \frac{P_jZ_i + P_iZ_j - Z_iZ_j\left( U_{j}^R - U_{i}^R \right)}{Z_i + Z_j},
\end{equation}
\begin{equation}
\label{RazpadU}
U^{*R}_{ij} = \frac{U_i^{R}Z_i + U_j^{R}Z_j - P_j + P_i}{Z_i + Z_j},
\end{equation}
where $U^R = \overrightarrow{U}\cdot\overrightarrow{r}_{ij}/\left| \overrightarrow{r}_{ij} \right|=\overrightarrow{U}\cdot\overrightarrow{e^R}$, $Z = \rho C$ is the acoustic impedance of the particle material.
To understand the reason for high numerical viscosity in the contact SPH method, let us consider the scheme given in Fig.~\ref{sph-viscosity-scheme}. It illustrates the mutual movement of two inviscid flows in opposite directions with velocities $\overrightarrow{U_i}$ and $\overrightarrow{U_j}$. Both liquids are in contact along the $x=0$ interface, the pressures are $P_{i}=P_{j}=0$. Figure~\ref{sph-viscosity-scheme}(a) shows the displacement of inviscid fluids, which evolution is guided by equations (\ref{Euler1})--(\ref{Grun}), while Fig.~\ref{sph-viscosity-scheme}(b) represent the evolution provided by the SPH method (\ref{SPHmass1})--(\ref{SPHenergy1}). Circles representing SPH-particles shows the flow configuration near the interface. The control surface connecting the centers of the particles is shown with the dashed line in both figures. At the time $t=0$ the control surface coincides with $x$-axis. The boundaries at $x=-L$ and $x=L$ are rigid walls without sticking, periodic boundary conditions are set along $y$ and $z$ axes.

According to Eq.~(\ref{Euler2}), there is no momentum exchange accross the interface between liquids when $P=0$. Therefore, the flow described by this equation  appears on both sides of the interface with constant velocities in opposite directions, as shown in Fig.~\ref{sph-viscosity-scheme}(a). This flow pattern is consistent with the interpretation of shear movement of inviscid liquids.

Eq.~(\ref{SPHmomentum1}) describes the central interaction of the SPH particles through the approximate Riemann problem solution~\eqref{RazpadP}, which leads to an exchange of momentum between particles $i$ and $j$ across the interface between liquids: even for $P_i=P_j=0$ Eq.~(\ref{RazpadP}) provides $P^*_{ij}\ne 0$. The presence of this pressure has a diffusive effect on the momentum of $i$ and $j$ particles moving in opposite directions, so that the flow profile is distorted as shown in Fig.~\ref{sph-viscosity-scheme}(b). This flow pattern is typical for viscous liquids: interface particles are slowing due to viscous friction. However, in contact SPH it appears due to the numerical viscosity of the method.

\section{Estimation of numerical viscosity in the contact SPH method}

The magnitude of the numerical viscosity can be estimated from the hypothesis that its effect on the fluid motion is equivalent to that of the physical viscosity. To test this hypothesis one may compare the velocity profile in the shear flow simulated according Eqs.~(\ref{SPHmass1})--(\ref{SPHenergy1}) and (\ref{Grun}) with the known analytical solution for the viscous diffusion~\cite{Lamb:1932}, which describes the velocity evolution $U_y(x,t)$ of all particles in the flow:
\begin{equation}
\label{PlateX0}
U_y(x,t)=U_{y}(x,0) \textrm{erf}\left(\frac{x}{\sqrt{4\nu t}}\right),
\end{equation}
where $\nu$ is the kinematic viscosity of the medium.

\subsection{\label{sec:shear-flow-setup}Problem statement}

Let us consider the motion of liquids in opposite directions along the flat interface as shown in
Fig.~\ref{sph-viscosity-scheme}(b). Initially, two samples of the liquid lead, shown in Fig.~\ref{shear-numeric-viscosity}(a), occupy a total volume of $800\,$nm$\times400\,$nm$\times40\,$nm. The flat interface between these flows passes vertically at $x=0$. At the initial moment of time $t=0$ both samples obtain the velocities $U_{y0}=\pm500\,$m/s equal in magnitude, but opposite in direction, parallel to the axis $y$, with $U_y>0$ at $x>0$ and $U_y<0$ at $x<0$. At the boundaries $x=-400\,$nm and $x=400\,$nm of the computational domain rigid walls with slip condition are set, while at the boundaries $y=-200\,$nm and $y=200\,$nm, $z=-40\,$nm and $z=40\,$nm periodic boundary conditions are applied. All particles have the same size $D_0=10\,$nm.

The linear size of a particle is determined by the dependence
\begin{equation}
\label{diamet}
D_i=\left(m_i/ \rho_i\right)^{1/d},
\end{equation}
where $d$ is the spatial dimension, and the smoothing distance $h_{ij}$ by the expression
\begin{equation}
\label{distance}
h_{ij}=\alpha\left(D_i+D_j\right),
\end{equation}
where $\alpha=0.5$. Two different smoothing kernels $W(r,h)$ are used: Wendland $C^2$~\cite{Wendland:1995} and the cubic B-spline~\cite{MonaghanBSpline}. The physical properties of real liquids are given in Table~\ref{TPhPM}.

\subsection{Numerical viscosity calculations}

\begin{figure}[t]
\centering
\includegraphics[width=1.0\textwidth]{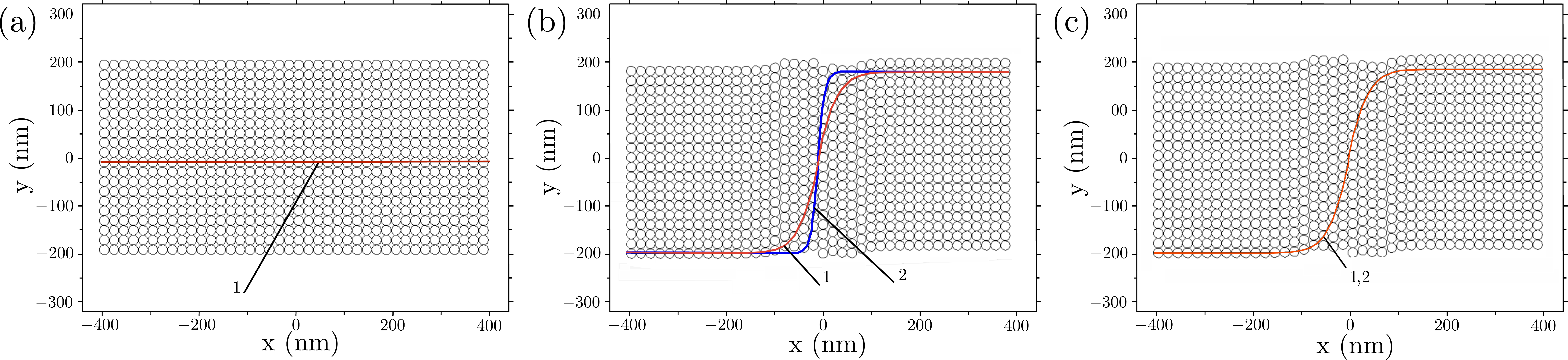}
\caption {\label{shear-numeric-viscosity}
(a) Initial position of SPH-particles at time $t=0$ and the control surface 1, connected in the process of viscous breakup evolution with the centers of the particles.
(b) Control surface 1 at time $t=0.19\,$ns and control surface 2 calculated by (\ref{PlateX0}) for liquid lead with dynamic viscosity $\eta=2.3\,$mPa$\,\cdot\,$s.
(c) Control surface 1 at time $t=0.19\,$s and control surface 2 calculated by (\ref{PlateX0}) for liquid lead with matched dynamic viscosity $\eta^{\mathrm{num}}=19.173\,$mPa$\,\cdot\,$s.
}
\end{figure}

The shear flow evolution results in the distortion of the control surface passing through the centers of the particles, which initially lie in the plane $y=-D_0/2$. At $t=0$ the control surface is a plane perpendicular to the materials interface as shown in Fig.~\ref{shear-numeric-viscosity}(a) by Line 1. Later, the control surface changes its shape according to the displacement of particles. For $t=0.19\,$ns the control surface (Line 1) is shown in Fig.~\ref{shear-numeric-viscosity}(b). Line 2 in Fig.~\ref{shear-numeric-viscosity}(b) is calculated using the relationship (\ref{PlateX0}) for liquid lead with the real dynamic viscosity of $\eta=2.3\,$mPa$\,\cdot\,$s. Thus, Line 2 predicts the control surface shape corresponding to the flow with viscous stresses of only physical origin. The discrepancy between the control surface shape obtained in SPH simulation (Line 1) and the theoretical one (Line 2), shown in Fig.~\ref{shear-numeric-viscosity}(b), is due to the presense of numerical viscosity in the contact SPH method, the origin of which is illustrated in Fig.~\ref{sph-viscosity-scheme}(b).

\begin{table}[h]
	\caption{\label{TPhPM}
	Values of physical $\eta$ \cite{PhysQuantities:1991} and numerical $\eta^{\mathrm{num}}$ viscosities for different materials. The numerical viscosity was estimated for the smoothing length parameter $\alpha=0.5$, particle size $D_0=10\,$nm, and two different smoothing kernels $W(r,h)$ (Wendland $C^2$/B-spline)}
	\centering
	\begin{tabular}{ |l|c|l|c|c| }
		\hline
		Material & $\rho_0$, kg/m$^3$  & $c_0$, m/s  & $\eta$, mPa$\cdot$s  & $\eta^{\mathrm{num}}$, mPa$\cdot$s   \\ \hline
		Lithium          &   534  &  4718   &   0.566    & 2.262/2.72   \\ \hline
		Benzene          &   880  &  1571.8 &   0.604    & 1.235/1.469  \\ \hline
		Paraffin         &   910  &  895.6  & 25-80      & 0.863/0.947  \\ \hline
		Water            &   1000 &  1483   & 1.004      & 1.353/1.58   \\ \hline
		Nitromethane     & 1128   & 1144.7  &  0.63      & 1.264/1.45  \\ \hline
		Glycerol         &   1250 &  1902.6 & 1480       & 2.074/2.524   \\ \hline
		Sodium chloride  & 2160   &  3496   &  1.03      & 6.672/8.145  \\ \hline
		Silica           &   2480 & 4181.18 & 10.0       & 9.29/11.206   \\ \hline
		Lithium fluoride & 2650   &  5132.2 &   ---      & 12.206/14.675  \\ \hline
		Aluminum         & 2688.9 &  5416   & 1.24       & 13.075/15.719  \\ \hline
		Corundum         & 3920   &  5779.8 &    ---     & 20.226/24.471  \\ \hline
		Titanium         & 4510   &  4924   &  4.42      & 20.047/23.969  \\ \hline
		Zinc             & 6577   &  3057   &  3.737     & 17.942/21.580  \\ \hline
		Iron             & 7880   &  4660   &  5.22      & 32.96/39.643 \\ \hline
		Nickel           & 8902   &  4999.9 &  5.01      & 40.165/48.015  \\ \hline
		Copper           & 8920   &  4118.5 &  3.92      & 32.852/39.702  \\ \hline
		Lead             & 11350  &  1931.7 &  2.3        & 19.173/23.278  \\ \hline
		Tantalum         & 16600  &  3481.5 &  8.5        & 51.034/62.332  \\ \hline
		Uranus           & 19100  & 3056.88 &  6.5        & 52.103/62.668  \\ \hline
		Tungsten         & 19350  &  3937.5 &  7.0        & 67.786/82.332  \\ \hline
	\end{tabular}
\end{table}

Let us denote the numerical dynamic viscosity $\eta^{\mathrm{num}}=\nu^{\mathrm{num}}\rho_0$. Using the expression~(\ref{PlateX0}) one may find the numerical value $\nu$ such that Line 2 would coincide with Line 1. The result for liquid lead is shown in Fig.~\ref{shear-numeric-viscosity}(c), which corresponds to the value $\nu=\eta/\rho_0$ at $\eta=19.173\,$mPa$\,\cdot\,$s used the expression (\ref{PlateX0}). From the Fig.~\ref{shear-numeric-viscosity}(c) it can be seen that the simulated and theoretical control surface shapes coincide. This means that the viscosity of numerical and physical origin affect the flow velocity in similar way. The value $\eta$ obtained with (\ref{PlateX0}) is the value of dynamic numerical viscosity $\eta^{\mathrm{num}}=\eta$ for this simulation. It should be noted that in this case, the numerical viscosity is almost an order of magnitude higher than the physical one.

\begin{figure}[t]
	\centering
	\includegraphics[width=1.0\linewidth]{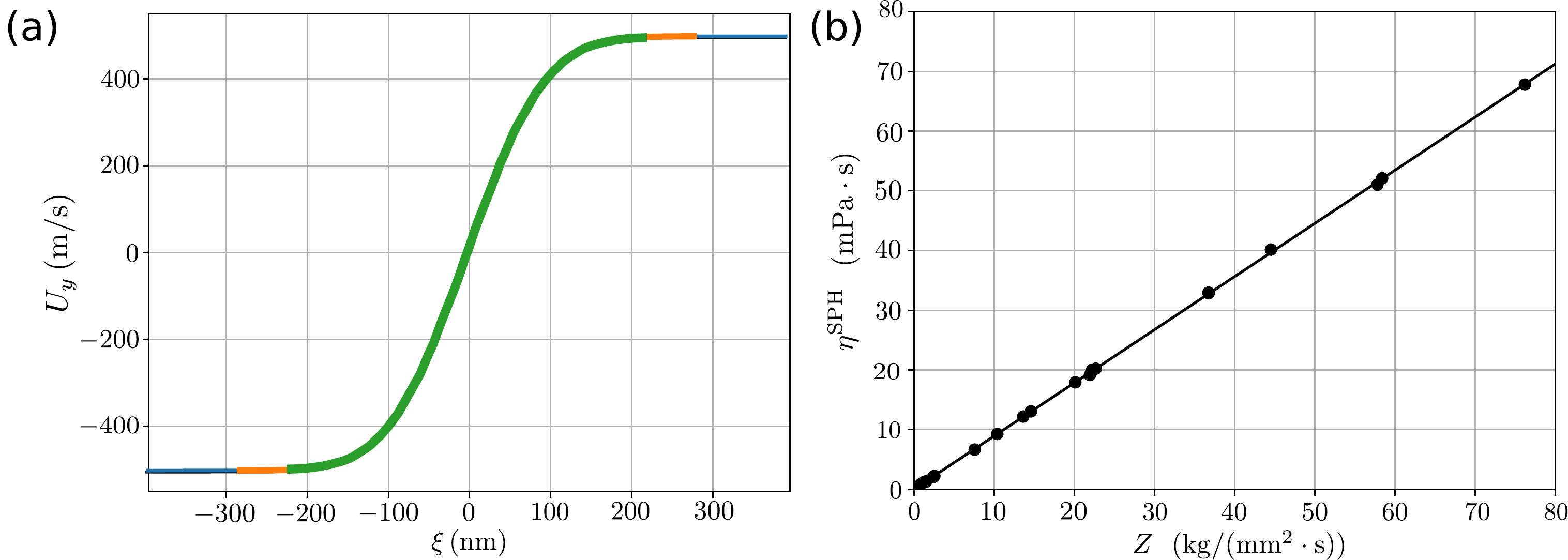}
	\caption{\label{self-similarity}
(a) Self-similarity of the velocity profile in the shear flow of liquid lead. The simulation is performed using SPH with the particle size $D_0 = 10\,$nm, the smoothing length parameter $\alpha = 0.5$, and Wendland C$^2$ smoothing kernel. The results are presented for time moments $t = 3\,$ns, $4.5\,$ns, and $5\,$ns for self-similar variable $\xi = x\sqrt{t_0/t}$. (b) Dependence of the numerical viscosity $\eta^{\mathrm{num}}$ on the acoustic impedance $\rho_0 c_0$ for the materials given in Table~\ref{TPhPM}.}
\end{figure}

To find out whether $\eta^{\mathrm{num}}$ retains its value during shear flow, the calculations  in the above formulation are performed for the time interval $[3,5]\,$ns. Figure~\ref{self-similarity}(a) shows that the velocity profiles as a function of $\xi = x\sqrt{t_0/t}$ remain constant over time, indicating that the resulting solutions are self-similar and, therefore, that the numerical viscosity is constant during the evolution of the flow.

The proposed approach can be applied to estimate the numerical viscosity for a number of materials with different values of density $\rho_0$, sound speed $c_0$, and dynamic viscosity $\eta$ of physical nature. The results are given in Table~\ref{TPhPM}. Figure~\ref{self-similarity}(b) shows the dependence of the numerical viscosity on the acoustic impedance of the material $Z=\rho_0 c_0$. It may be clearly seen that the value of the numerical viscosity is linearly proportional to the acoustic impedance of the material. The particle size in our calculations is chosen to be $D_0=10\,$nm, so that the values of the physical viscosity of the medium and the numerical viscosity of the method are approximately of the same order, which allows us to observe and compare the analytical and numerical shear flow velocity profiles.

\begin{figure}[t]
	\centering
	\includegraphics[width=1.0\linewidth]{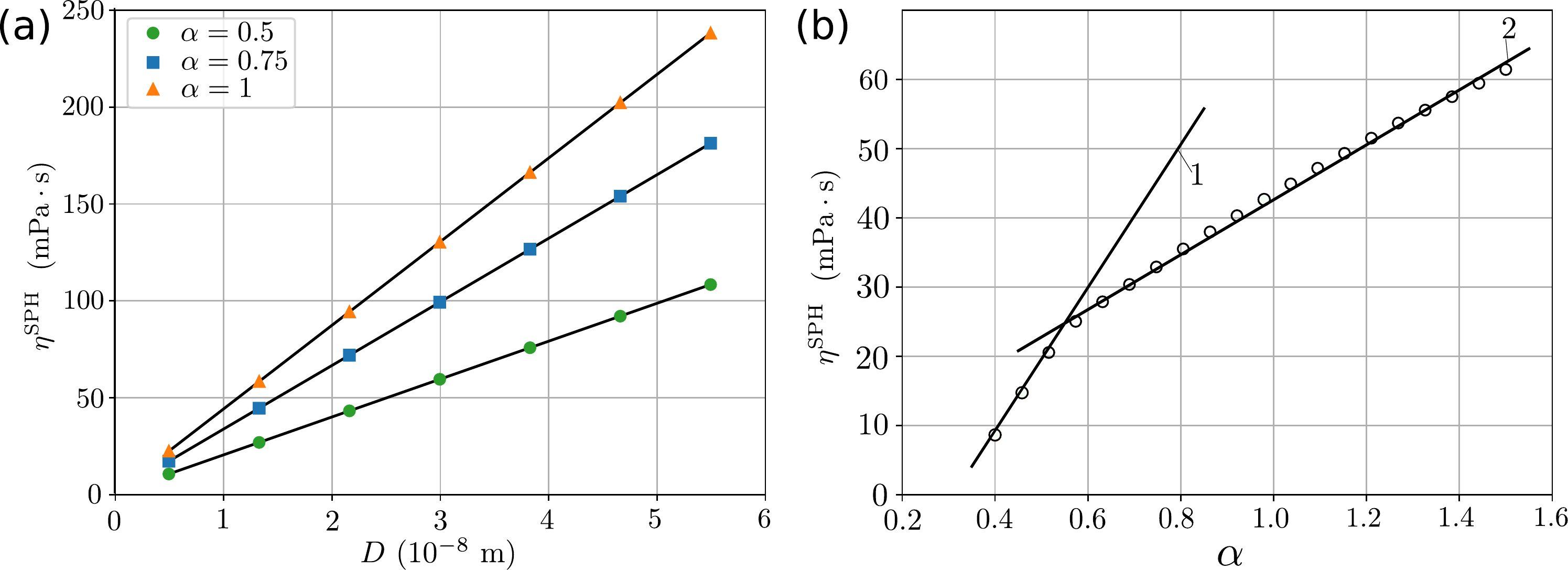}
	\caption{\label{viscosity-size-smoothing}
(a) Dependence of the numerical viscosity $\eta^{\mathrm{num}}$ on the particle size at different values of $\alpha=0.5,\,0.75,\,1.0$. (b) Dependence of the numerical viscosity $\eta^{\mathrm{num}}$ on the smoothing length parameter $\alpha$ according to~\eqref{distance}.
}
\end{figure}

It is worth noting, that similar study may be performed with different particle sizes and smoothing length parameters. Figure~\ref{viscosity-size-smoothing}(a) shows that numerical viscosity linearly depends on the particle size $D_0$ for various values of smoothing length parameter $\alpha$. However, the dependence of $\eta^{\mathrm{num}}$ on the smoothing length parameter $\alpha$ with the fixed particle size is more difficult. Figure~\ref{viscosity-size-smoothing}(b) shows that it may be approximated well with a combination of two linear functions. Thus, the numerical viscosity depends on the listed parameters as
\begin{equation}
\label{eta}
\eta^{\mathrm{num}}\thicksim \alpha D_0 Z.
\end{equation}
Calculations performed with various initial velocity $U_{y0}$ in the range $[50,500]\,$m/s with step $50\,$m/s show that the dynamic numerical viscosity of the method is almost independent of the flow velocity, and Eq.~\eqref{eta} is still valid.

The expression~(\ref{eta}) shows that it is possible to control the value of the numerical viscosity by adjusting the particle size and the parameter $\alpha$, which has a lower limit of $0.5$ and usually does not exceed the value of $2$. The acoustic impedance of the simulated fluids is determined by physical properties of realistic materials. Thus, simulation of realistic flows is possible when the particle sizes are small: one may ignore the effect of numerical viscosity by choosing $D\lesssim 1\,$nm and account only for physical viscosity. However, such restriction is unacceptable for most problems due to computational costs or computer memory limitations. In fact, it is the scale of molecular dynamics simulations, which may provide more realistic description of materials with appropriate interatomic potentials. For larger scales, the numerical viscosity control becomes an extremely urgent problem.

\section{Numerical viscosity handling in the SPH contact method}

It is shown above that the numerical and physical viscosity demonstrate identical behavior when simulating the shear flow of compressible fluids. Therefore, to control the numerical viscosity $\eta^{\mathrm{num}}=\nu^{\mathrm{num}}\rho_0$ it is possible to construct a special viscous stress tensor which magnitude may be adjusted in the equations for calculating viscous flows. To describe the flow of a viscous compressible fluid, equations (\ref{Euler2}) and (\ref{Euler3}) are extended as follows:
\begin{equation}
\label{momentum}
\rho  \frac{d\overrightarrow{U}}{dt} = \nabla \cdot \mathbf{\Pi},
\end{equation}
\begin{equation}
\label{energy}
\rho \frac{dE}{dt} = \nabla \cdot \left(\mathbf{\Pi}\cdot \overrightarrow{U}\right),
\end{equation}
where the stress tensor $\mathbf{\Pi}$ is defined as:
\begin{equation}
\label{tens_full}
\Pi^{\alpha\beta}=-P\delta^{\alpha \beta}+\tau^{\alpha\beta}.
\end{equation}
Here $ \tau^{\alpha\beta}$ is viscous stress tensor-deviator, $\delta^{\alpha \beta}$
is the identity tensor, $\alpha=x,y,z$, and $\beta=x,y,z$.
The components of the viscous stress tensor for Newtonian fluids are expressed as:
\begin{equation}
\label{tens_viscosity}
\tau^{\alpha\beta}=\eta \left[ \left(\frac{\partial U^\alpha}{\partial r^{\beta}}
+\frac{\partial U^\beta}{\partial r^{\alpha}}\right)
-\frac{2}{3} \left( \nabla \cdot \overrightarrow{U}   \right)
\delta^{\alpha \beta} \right],
\end{equation}
where $\eta$ is the dynamic viscosity coefficient of the fluid, $U^\alpha$ and $U^\beta$ are components of the velocity vector $\overrightarrow{U}$. The stress tensor $\bf\Pi$ in a viscous compressible fluid can be expressed as the sum of the isotropic tensor and the deviator:
\begin{equation}
\label{Sigma1}
\bf\Pi=-\bf P+\bf T = \left(
\begin{array}{lcr}
-p  &  0   &  0  \\
0   &  -p  &  0  \\
0   &  0   &  -p  \\
\end{array}
\right)
+
\left(
\begin{array}{lcr}
\tau_{xx}  &  \tau_{xy}  &  \tau_{xz}  \\
\tau_{yx}  &  \tau_{yy}  &  \tau_{yz}  \\
\tau_{zx}  &  \tau_{zy}  &  \tau_{zz}  \\
\end{array}
\right).
\end{equation}

Above we noted, that in a discrete medium consisting of SPH particles in addition to the stresses $\bf\Pi$ (\ref{Sigma1}) Riemann problem solutions at interparticle contacts provides viscous stresses of not physical, but numerical origin. Formally, this means that instead of the tensor $\bf\Pi$ in a discrete medium some other tensor $\bf\Pi^{'}$ is calculated which contains (besides physical ones) also the components of numerical origin which causes the numerical viscosity $\eta^{\mathrm{num}}$. Let us denote this tensor as $\bf\Pi^{\mathrm{num}}$. Then the viscous stress tensor for the discrete SPH medium $\bf\Pi^{'}$ can be represented as the sum of the viscous stress tensor of physical origin $\bf\Pi$ and the deviator tensor of numerical origin $\bf T^{\mathrm{num}}$:
\begin{equation}
\label{Sigma11}
\bf\Pi^{'}=\bf\Pi +\bf T^{\mathrm{num}} =-\bf P+\bf T + \bf T^{\mathrm{num}} =-\bf P+\bf T + \left(
\begin{array}{lcr}
\tau_{xx}^{\mathrm{num}}  &  \tau_{xy}^{\mathrm{num}}  &  \tau_{xz}^{\mathrm{num}}  \\
\tau_{yx}^{\mathrm{num}}  &  \tau_{yy}^{\mathrm{num}}  &  \tau_{yz}^{\mathrm{num}}  \\
\tau_{zx}^{\mathrm{num}}  &  \tau_{zy}^{\mathrm{num}}  &  \tau_{zz}^{\mathrm{num}}  \\
\end{array}
\right),
\end{equation}
where the tensor components $\bf T^{\mathrm{num}}$ may be expressed using the dynamic numerical viscosity coefficient $\eta^{\mathrm{num}}$ and physical quantities (such as velocity) which corresponds the viscous media:
\begin{align}
\label{tauxx}
\tau_{xx}^{\mathrm{num}}=\frac{2}{3}\eta^{\mathrm{num}}\left(2\frac{\partial U_x}{\partial x}-\frac{\partial U_y}{\partial y}-\frac{\partial U_z}{\partial z} \right), \\
\label{tauyy}
\tau_{yy}^{\mathrm{num}}=\frac{2}{3}\eta^{\mathrm{num}}\left(2\frac{\partial U_y}{\partial y}-\frac{\partial U_x}{\partial x}-\frac{\partial U_z}{\partial z} \right),  \\
\label{tauzz}
\tau_{zz}^{\mathrm{num}}=\frac{2}{3}\eta^{\mathrm{num}}\left(2\frac{\partial U_z}{\partial z}-\frac{\partial U_x}{\partial x}-\frac{\partial U_y}{\partial y}   \right), \\
\tau_{xy}^{\mathrm{num}}=\eta^{\mathrm{num}}\left(\frac{\partial U_x}{\partial y}+\frac{\partial U_y}{\partial x}\right)=\tau_{yx}^{\mathrm{num}}, \\
\tau_{xz}^{\mathrm{num}}=\eta^{\mathrm{num}}\left(\frac{\partial U_z}{\partial x}+\frac{\partial U_x}{\partial z}\right)=\tau_{zx}^{\mathrm{num}}, \\
\tau_{yz}^{\mathrm{num}}=\eta^{\mathrm{num}}\left(\frac{\partial U_y}{\partial z}+\frac{\partial U_z}{\partial y}\right)=\tau_{zy}^{\mathrm{num}}.
\end{align}
It follows from the expression (\ref{Sigma11}) that modeling of viscous flow by the contact SPH method with the numerical viscosity reduced by $\zeta\cdot 100\,\%$ can be done by a simple procedure of subtracting from $\bf\Pi^{'}$ the viscous stress of numerical origin
\begin{equation}
\label{Sigma13}
\bf\Pi = \bf\Pi^{'} - \zeta\bf T^{\mathrm{num}}.
\end{equation}
According to the equations (\ref{tauxx})--(\ref{tauzz}), it is sufficient to use the value $\zeta\eta^{\mathrm{num}}$ where $0\leqslant\zeta\leqslant1$ as the numerical viscosity. Let us consider the procedure (\ref{Sigma13}) in detail.

\begin{figure}[t]
    \centering
	\includegraphics[width=1.0\linewidth]{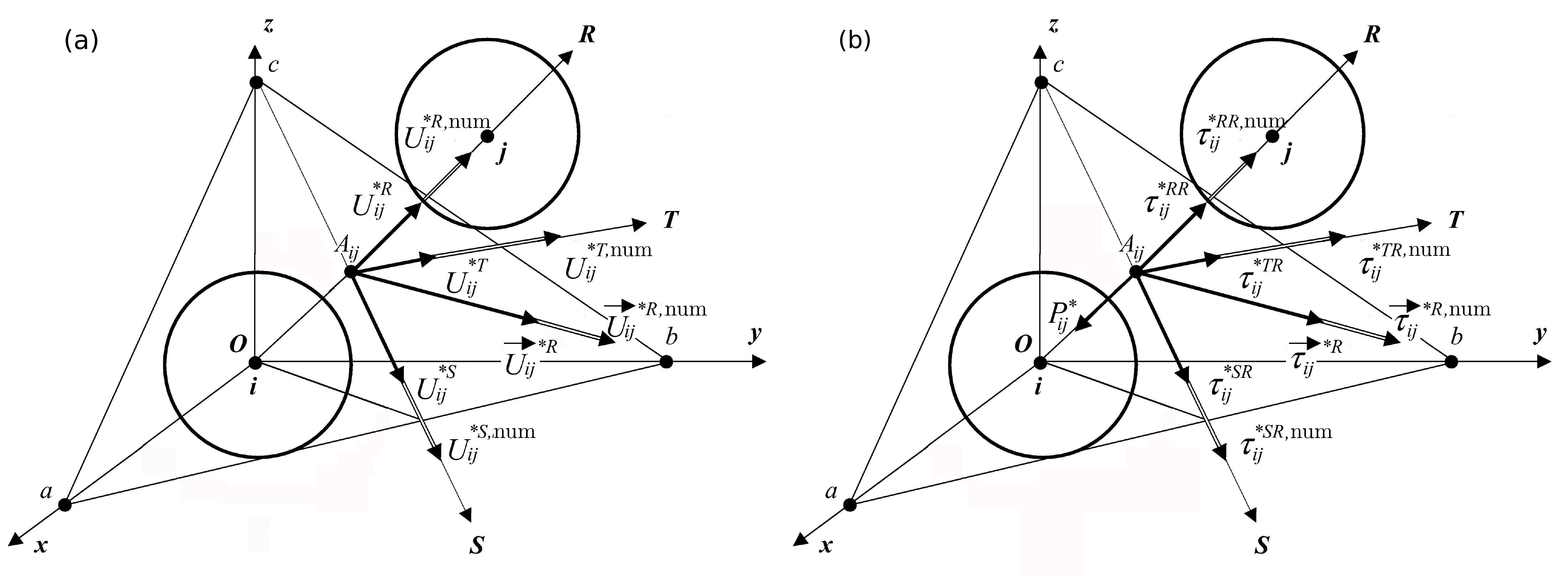}
	\caption {\label{viscous-contact}
	(a) Velocity components at inter-particle contact surface during viscosity tensor evaluation. (b) Stress components at inter-particle contact surface during viscosity tensor evaluation.}
\end{figure}

Figure~\ref{viscous-contact}(a) shows the velocities in the $RST$ coordinate system during the evaluation of viscous stresses at the contact between the main particle $i$ and the neighbor particle $j$. This system is related to the tangent plane of the particles $abc$, which passes through the point $A_{ij}$. The point $A_{ij}$ divides the distance between the particles into parts proportional to the linear sizes of the particles.
As was postulated in \cite{Parshikov:JCP:2002}, the $abc$ contact plane of particles $i$ and $j$ in a discrete {\sl SPH}-medium is equivalent to the contact surface in a continuous medium. Therefore, we can construct the Riemann solution for the discontinuity along the axes $R,S,T$ and calculate the following velocity components $U_{ij}^{*RR}$, $U_{ij}^{*RS}$ and $U_{ij}^{*RT}$ at the point $A_{ij}$. The velocity components of the medium at the contact point \cite{German:1973}:
\begin{equation}
\label{VelVisR2}
U^{\gamma*}_{ij}= \frac{\eta_i U^{\gamma}_i+\eta_j U^{\gamma}_j\sqrt{\nu_i/\nu_j} }
{\eta_i+\eta_j \sqrt{\nu_i/\nu_j}} , \quad \gamma = R,S,T.
\end{equation}
Figure~\ref{viscous-contact}(b) shows the scheme of the viscous stresses at the contact between the main particle $i$ and the neighbour particle $j$ in the coordinate system $RST$.

Assuming $\sqrt{\eta_j \rho_i}/ \sqrt{\eta_i \rho_j}=\varkappa$, one may write the expressions for the components of the viscous stress vector at the contact point $A_{ij}$:
\begin{equation}
\label{TauRi}
\tau^{RR*}_i= \frac{4}{3}\eta_i\frac{U^{R*}_{ij}-U^R_i}{\Delta r_i},
\quad \text{где} \quad \Delta r_i= \frac{|\overrightarrow{r_j}-\overrightarrow{r_i}|}{1+\varkappa}=
\frac{\sqrt{\nu_i/\rho_j}|\overrightarrow{r_j}-\overrightarrow{r_i}|}{1+\sqrt{\nu_i/\rho_j}},
\end{equation}
\begin{equation}
\label{TauRj}
\tau^{RR*}_j= \frac{4}{3}\eta_j\frac{U^R_j-U^{R*}_{ij}}{\Delta r_j},
\quad \text{где} \quad \Delta r_j= \frac{\varkappa|\overrightarrow{r_j}-\overrightarrow{r_i}|}
{1+\varkappa}=
\frac{|\overrightarrow{r_j}-\overrightarrow{r_i}|}{1+\sqrt{\nu_i/\rho_j}}.
\end{equation}
Combining the formulas above, we obtain:
\begin{equation}
\label{TauRR}
\tau^{RR*}_{ij}= \tau^{RR*}_i= \tau^{RR*}_j=  \frac{4}{3}\eta_i\eta_j \frac{1+\sqrt{\nu_i / \nu_j}}{\eta_i+\eta_j\sqrt{\nu_i/\nu_j}}
\frac{U^R_j-U^R_i}{|\overrightarrow{r_j}-\overrightarrow{r_i}|},
\end{equation}
\begin{equation}
\label{TauSR}
\tau^{SR*}_{ij}= \tau^{SR*}_i= \tau^{SR*}_j=  \eta_i\eta_j \frac{1+\sqrt{\nu_i / \nu_j}}{\eta_i+\eta_j\sqrt{\nu_i/\nu_j}}
\frac{U^S_j-U^S_i}{|\overrightarrow{r_j}-\overrightarrow{r_i}|},
\end{equation}
\begin{equation}
\label{TauTR}
\tau^{TR*}_{ij}= \tau^{TR*}_i= \tau^{TR*}_j=  \eta_i\eta_j \frac{1+\sqrt{\nu_i / \nu_j}}{\eta_i+\eta_j\sqrt{\nu_i/\nu_j}}
\frac{U^T_j-U^T_i}{|\overrightarrow{r_j}-\overrightarrow{r_i}|}.
\end{equation}

All the necessary variables for the construction of the desired contact SPH equations for the viscous medium are now defined. It is reasonable to represent the continuity equation (\ref{SPHmass1}) as an equation for the strain rate by dividing the right and left parts by $\rho_i$:
\begin{equation}
\label{mynereps}
\frac{ d \varepsilon_i}{dt}= \dot\varepsilon_i =- 2\sum_{j}\frac{ m_j }{\rho_j} \left[
\left({U_i}^R-{U_{ij}}^{*R}+{U_{ij}}^{*R,\mathrm{num}}\right)\overrightarrow{e^R}\right]\cdot \nabla_i W_{ij}.
\end{equation}
Then the following expression can be applied to calculate the density evolution using local integration over the time step $\Delta t$ in the transition from time layer $n$ to layer $n+1$:
\begin{equation}
\rho_i^{n+1}=\rho_i^{n}\exp\left(-\Delta t \dot\varepsilon_i\right).
\end{equation}

The equation of motion in the SPH-form, which takes into account the viscous stresses, takes the form:
\begin{equation}
\label{myimpV}
\frac{ d\overrightarrow{U_i}}{dt}=-
2\sum_{j}\frac{ m_j }{\rho_j \rho_i}\left[\left( P_{ij}^*l^{R\alpha}-\tau^{\gamma R*}_{ij}l^{\gamma
	\alpha}+\tau^{\gamma R*,\mathrm{num}}_{ij}l^{\gamma
	\alpha}\right)\overrightarrow{e^\alpha}\right]\overrightarrow{e^R}\cdot \nabla_i W_{ij},
\end{equation}
where $\overrightarrow\tau^R=\tau^{\alpha R}\overrightarrow{e^\alpha}=\tau^{\gamma R}l^{\gamma
	\alpha}\overrightarrow{e^\alpha}$, $\gamma = R,S,T$, $\alpha = x,y,z$, and $l^{\gamma \alpha}$ are directional cosines.
The energy equation taking into account viscous stresses takes the form:
\begin{equation}
\label{myencV}
\frac{ dE_i}{dt}= -2\sum_{j}\frac{ m_j } {\rho_j\rho_i}
\Bigl[P_{ij}^*\left(U_{ij}^{*R} - U_{ij}^{*R,\mathrm{num}}\right) - \left(\tau^{\gamma R*}_{ij}-\tau^{\gamma R*,\mathrm{num}}_{ij}\right) \left(U_{ij}^{*\gamma}-U_{ij}^{*\gamma,\mathrm{num}}\right) \Bigr]\overrightarrow{e^R}\cdot \nabla_i W_{ij}.
\end{equation}

All values with upper index ``num'' in Eqs.~(\ref{mynereps}),(\ref{myimpV}) and (\ref{myencV})  are calculated by dependences (\ref{VelVisR2}) and (\ref{TauRR})--(\ref{TauTR}), where instead of values $\eta$ and $\nu$, corresponding to physical viscosity, their numerical equivalents $\zeta\eta^{\mathrm{num}}$ and $\zeta\nu^{\mathrm{num}}$ are substituted. The system of contact SPH equations (\ref{mynereps}),(\ref{myimpV}),(\ref{myencV}) is closed with the equation of state (\ref{Grun}) and provides a complete description for the evolution of unsteady compressible viscous flows.

\section{Examples of numerical viscosity adjusting in fluid dynamics simulations}

To test the developed method of numerical viscosity handling, simulations of viscous fluid flows are performed and compared with analytical solutions. Hereafter, the Wendland $C^2$ kernel is used. The numerical viscosity (\ref{eta}) for particular material characterized by $Z = \rho_0 c_0$ may be approximated using the following expression:
\begin{equation}
\label{eq:numerical-viscosity-approximation}
\eta^{\mathrm{num}} =
\begin{cases}
D_0Z(0.47142\alpha-0.14629), ~~ \alpha \leq 0.55
\\
D_0Z(0.18093\alpha+0.01347),  ~~ \alpha > 0.55
\end{cases}
\end{equation}
where $D_0$ is given in meters, and $Z$ in kg/(m$^2$s). The viscosity in our simulations is reduced to a fraction $\zeta$ of this full numerical viscosity as $\zeta\eta^{\mathrm{num}}$, where $\zeta \in [0, 1]$.

\subsection{\label{sec:shear-identical}Momentum diffusion in viscous shear flow of identical fluids}

\begin{figure}[t]
\centering
\includegraphics[width=1.0\linewidth]{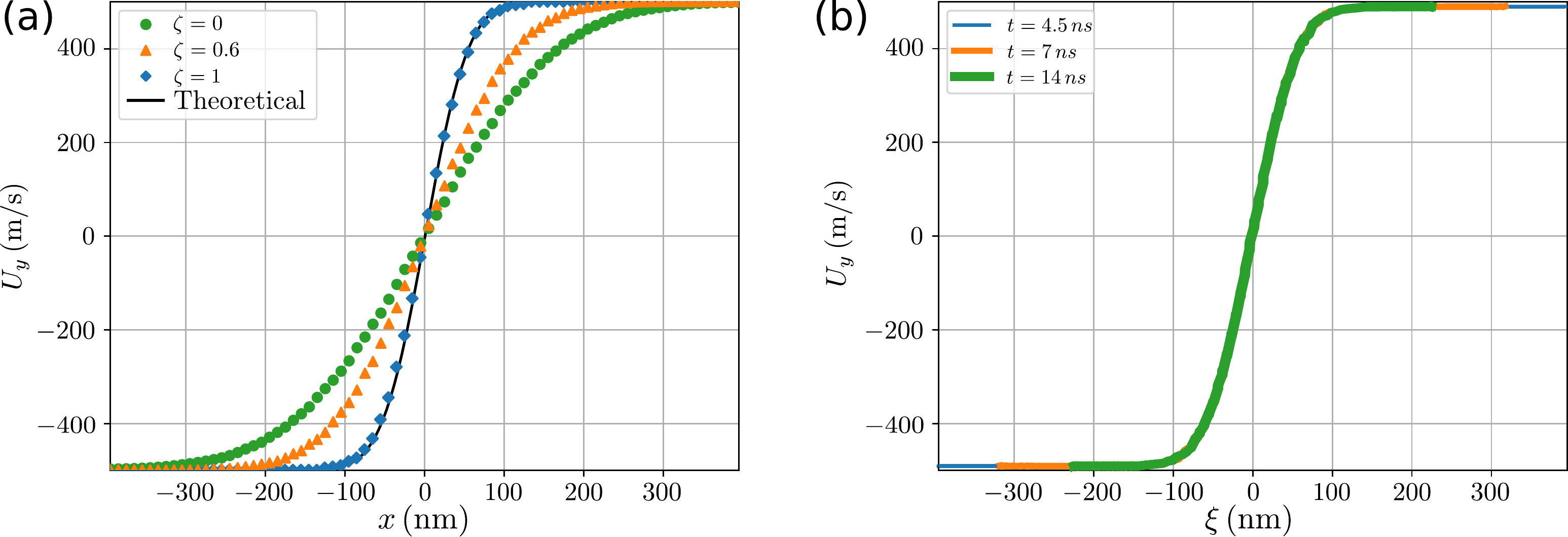}
\caption{\label{shear-viscosity-correction-automodel}
(a) Viscous momentum diffusion in liquid lead flows moving in opposite directions at $t = 4.5\,$ns for different values of $\zeta\eta^{\mathrm{num}}$. (b) Self-similarity in viscous momentum diffusion of liquid lead flows moving in opposite directions at $\zeta=1$ for time moments $t = 4.5\,$ns, $7\,$ns, $14\,$ns.}
\end{figure}

The momentum diffusion at the interface of opposite flows of liquid lead at the melting temperature is considered. The problem setup is similar to that described in Sec.~\ref{sec:shear-flow-setup}. The properties of the liquid lead are given in Table~\ref{TPhPM}. The calculated velocity profiles $U_y(x)$ for the time moment $t=4.5\,$ns at various values of $\zeta$ are shown in Fig.~\ref{shear-viscosity-correction-automodel}(a). As can be seen, the difference between the calculated velocity profile and its theoretical prediction~(\ref{PlateX0}) decreases with $\zeta$ growth. At $\zeta=1$, which corresponds to a complete exclusion of the numerical viscosity, the calculated velocity profile coincides with theory~(\ref{PlateX0}).

Figure~\ref{shear-viscosity-correction-automodel}(b) shows the dependence of fluid velocity on coordinate $\xi = x\sqrt{t_0/t}$ at $t = 4.5\,$ns, $7\,$ns, $14\,$ns, $t_0 = 4.5\,$ns and $\zeta=1$. It shows that self-similarity of the momentum diffusion is preserved in time as the flow continues.

\subsection{Momentum diffusion in viscous shear flow of different fluids}

The problem is similar to the previous one, but performed for different fluids propagating in opposite directions. The dimensions of the computational domain are also changed: both liquids (paraffin and glycerol) occupy a total volume of $600\,\mu$m$\times700\,\mu$m$\times80\,\mu$m with particle size $D_0=10\,\mu$m.
The region $-300\,\mu$m<$x$<$0$ is occupied by paraffin, while glycerol is placed within $0\,\mu$m<$x$<$300\,\mu$m.

\begin{figure}[t]
\centering
\includegraphics[width=1.0\linewidth]{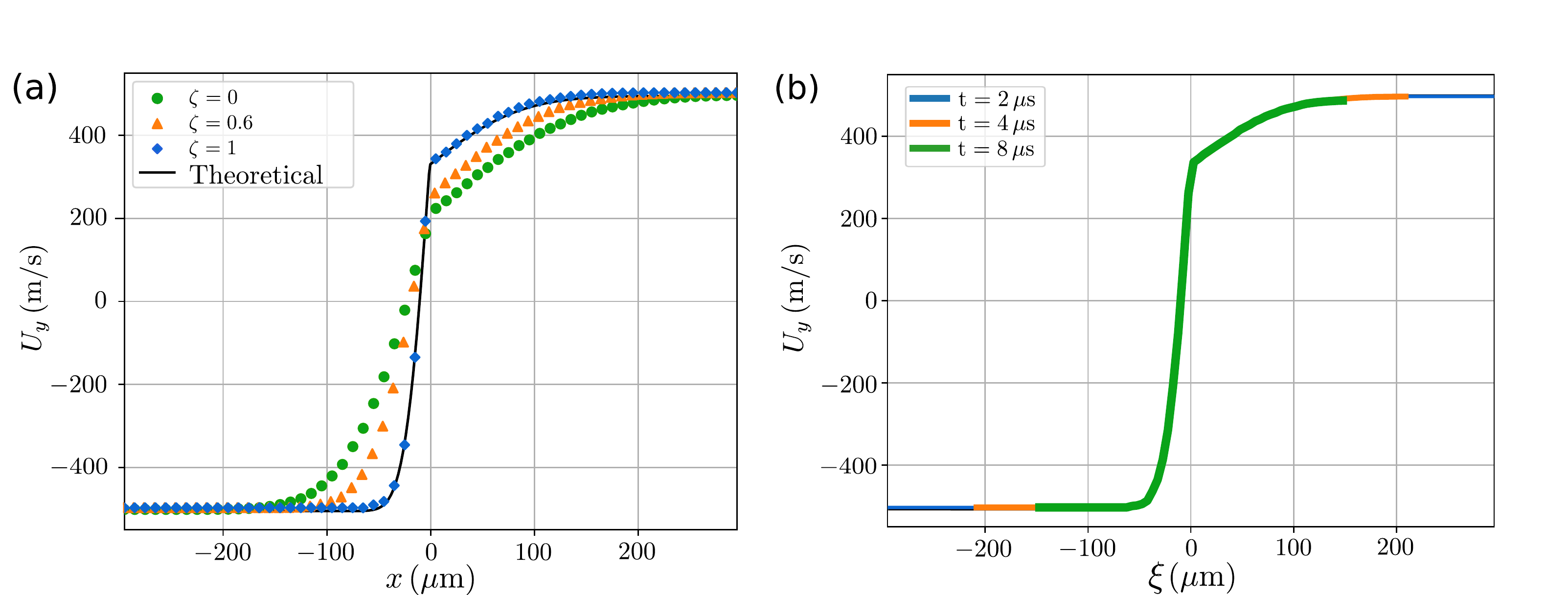}
\caption {\label{shear-viscosity-correction-automodel-paraffin-glycerol}
(a) Viscous momentum diffusion in liquid paraffin (left) and glycerin (right) flows moving in opposite directions at $t = 2\,\mu$s for different values of $\zeta\eta^{\mathrm{num}}$. (b) Self-similarity in viscous momentum diffusion at the interface of liquid paraffin (left) and glycerin (right) flows for time moments $t = 2\,\mu$s, $4\,\mu$s, $8\,\mu$s
}	
\end{figure}

Figure~\ref{shear-viscosity-correction-automodel-paraffin-glycerol}(a) shows the velocity profile $U_y$ which is formed accross the interface of two liquids. One may notice the effect of using Eq.~\eqref{VelVisR2} for evaluating the velocity at the interface. The increase in viscosity correction magnitude $\zeta$ to 1 allows to reach the desired theoretical profile.

Figure \ref{shear-viscosity-correction-automodel-paraffin-glycerol}(b) shows the velocity profiles at $t = 2\,\mu$s, $4\,\mu$s, $8\,\mu$s, $t_0=2\,\mu$s with maximum correction $\zeta = 1$ in self-similar manner using the coordinate $\xi = x\sqrt{t_0/t}$. One may notice, that velocity profiles are persistent in time and correspond the theoretical one.

The agreement of simulated velocity profiles with the theoretical solution demonstrate the applicability of our approach for viscosity adjustment: it does not lead to violation of the self-similarity and is suitable for calculating viscous multi-material flows.

\subsection{Falling ball test simulation}

\begin{figure}[t]
    \centering
    \includegraphics[width=1.0\linewidth]{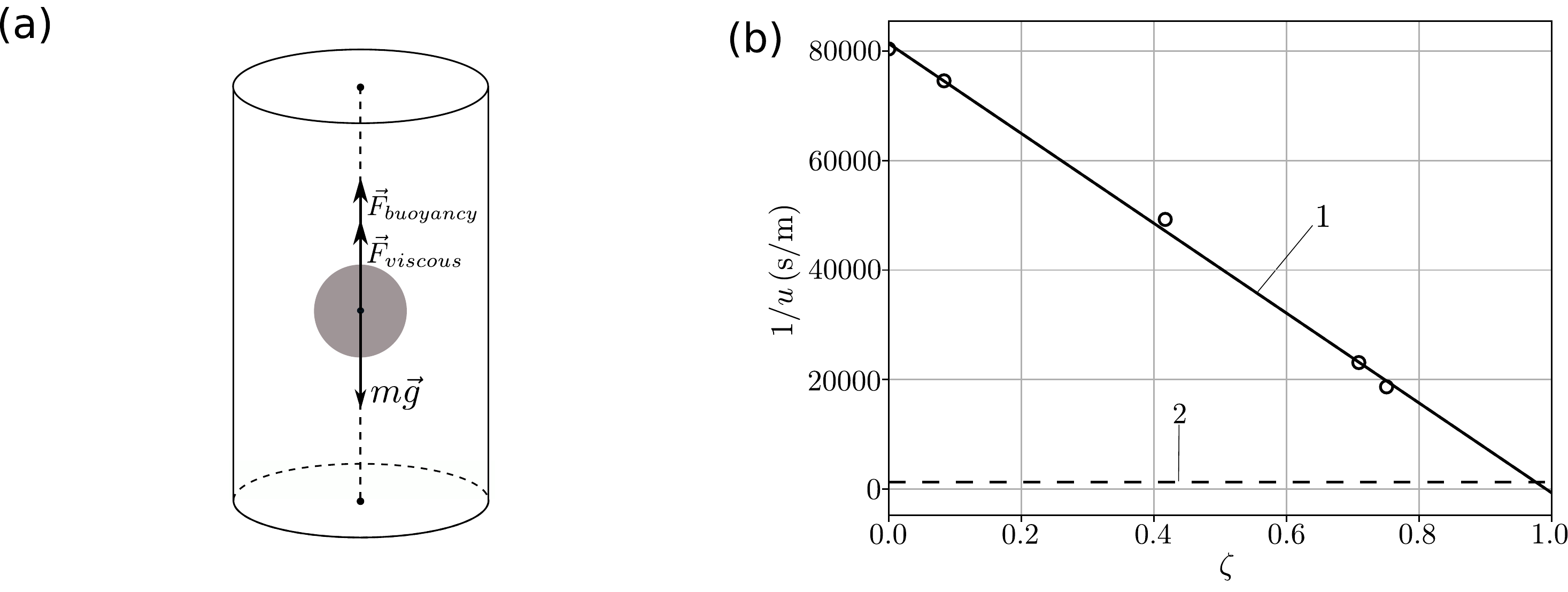}
    \caption {\label{stokes-problem-setup}
    (a) Scheme of the Stokes method for measuring liquid viscosity.
 (b) The inverse terminal velocity of an aluminum ball falling in water. Line 1 approximates the SPH calculations at $D = 0.5\,\mu$m and $\alpha = 0.5$. Line 2 corresponds to the inverse terminal velocity of the ball calculated using the dependence (\ref{Stokes}). }
\end{figure}

The Stokes method for measuring fluid viscosity is based on the experimental tracking of the ball velocity $u$ moving in a cylinder filled with liquid of lower density under the gravity force $g$. Due to the viscosity $\eta$, such ball reaches terminal velocity $u$, which is expressed with the following relation:
\begin{equation}
\label{Stokes}
u= \frac{2r^2g(\rho-\rho_0)}{9\eta\lambda},
\end{equation}
where $r$ is the radius of the ball, $\rho$ is the material density of the ball, $\rho_0$ is the fluid density, and $\lambda$ takes into account the effect of the vessel walls~\cite{Richou:EPJ:2003}. The ball reaches the terminal velocity $u$ at the moment, when gravitational force $m \overrightarrow g$ is equilibrated by the viscous friction and buoyancy force, as shown in Fig.~\ref{stokes-problem-setup}(a).

SPH simulation of such test is quite complicated compared to the previous ones. First, it is performed in three dimensions, so that the ball of size $r=35\,\mu$m consists of 1436644 particles with the size $D = 0.5\,\mu$m, while the cylindrical vessel of $70\,\mu$m radius and $364\,\mu$m height consists of 46774232 particles. The ratio of the ball radius to the cylinder radius is $0.5$, so that according to Ref.~\cite{Richou:EPJ:2003} $\lambda = 5.9638$. Second, the boundary condition setup between the ball and surrounding liquid should be properly defined: the hard ball may be considered as ``extremely viscous'' compared to liquid, so that numerical viscosity correction is not applied to the ball's particles. As a result, due to such numerical viscosity liquid particles also ``stick'' to the ball's surface, which is similar to real physical process. As a result, viscous friction appear at the interface between ``sticked'' to the ball particles of liquid and surrounding ones. The correction to numerical viscosity tensor is then applied only for liquid particles. The displacement of the ball until the terminal velocity achieved does not exceed $5\,$nm, so the deformation of the ball can be neglected, which justifies the use of the dependence~(\ref{Stokes}). Finally, at cylinder boundaries rigid walls are used.

Figure~\ref{stokes-problem-setup}(b) shows the dependence of the inverse terminal velocity of the aluminum ball falling in water on the numerical viscosity correction factor $\zeta$. Theoretical prediction of terminal velocity is given by Line 2 according to Eq.~(\ref{Stokes}). It is worth noting, that simulations at $\zeta=1$ (complete exclusion of the numerical viscosity of water) results in instabilities development, which may be due to unusual boundary conditions applied. Nevertheless, the successfull simulations at various $\zeta$ demonstrate the trend of terminal velocity convergence to the theoretical value.

\subsection{Consistency of viscosity adjustement method for 2D and 3D viscous flows}

\begin{figure}[t]
\centering
\includegraphics[width=1.0\linewidth]{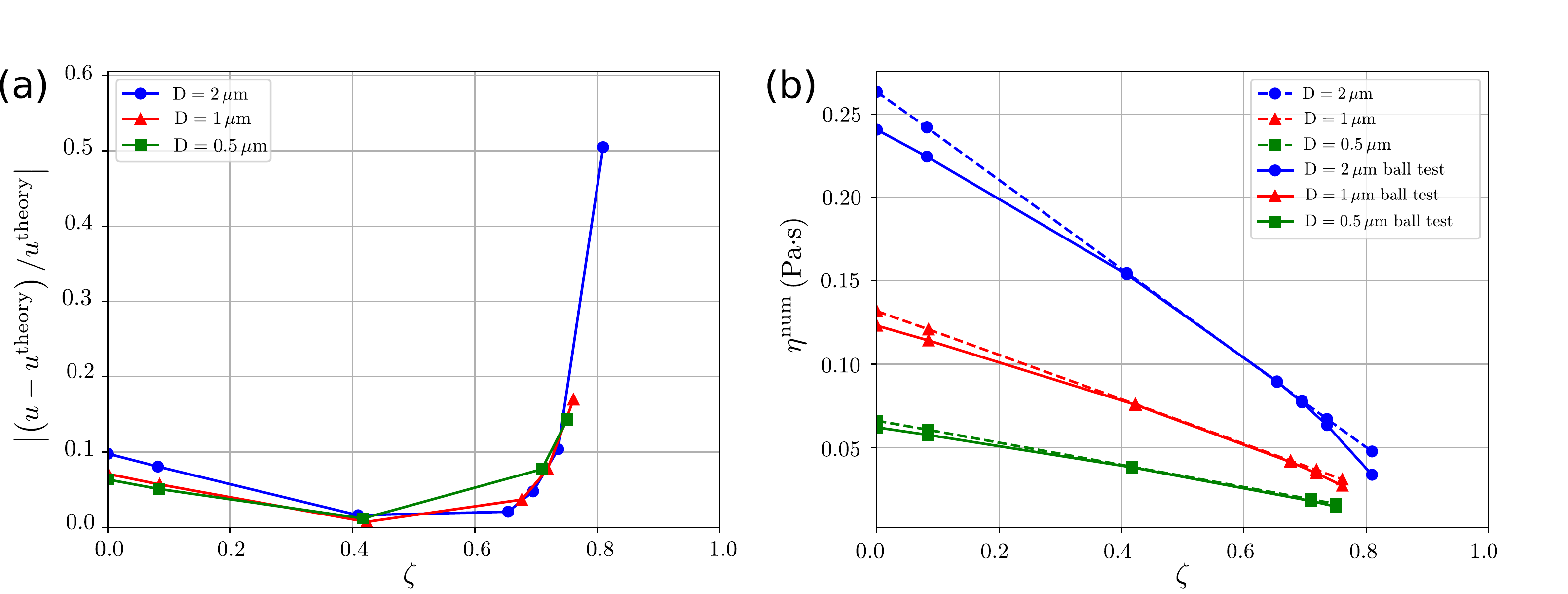}
\caption {\label{stokes-viscosity-comparison}
(a) Relative deviation of the ball's terminal velocity $u$ in calculations with different spatial discretization from the theoretical value $u^{\mathrm{theory}}$, obtained using the numerical viscosity in corresponding shear tests and substituted to Eq.~\eqref{Stokes} depending on the correction factor $\zeta$. (b) Comparison of the values $\nu^{\mathrm{num}}$ obtained by the Stokes method with those obtained in the corresponding shear tests.
}
\end{figure}

One may notice that the proposed algorithm for numerical viscosity evaluation is based on two-dimensional shear test providing quite specific dependencies which are inapplicable in real-life problems. To demonstrate the consistency of the numerical viscosity effect in various problems, here the falling ball test is used to measure numerical viscosity. Simulations are performed with different particle size $D = 0.5$, $1$, $2\,\mu$m which correspond to $35$, $70$ and $140$ particles per ball diameter. For consistency, the same spatial resolution is used in shear test with water flows moving in opposite directions as described in Sec.~\ref{sec:shear-identical}.

The analysis of spatial resolution effect on terminal velocity is shown in Fig.~\ref{stokes-viscosity-comparison}(a). Here the velocity $u^{\mathrm{theory}}$ is evaluated using Eq.~\eqref{Stokes} with the numerical viscosity obtained in the corresponding shear test (see also Table~\ref{TPhPM}). One may notice, that at moderate correction factors $\zeta \sim 0.5$ the best agreement is reached. The corresponding numerical viscosity is given in Fig.~\ref{stokes-viscosity-comparison}(b). One may clearly see that better spatial resolution (smaller particles) results in lower numerical viscosity. With the increase of correction factor $\zeta$ the numerical viscosity reduces. Above all, the consistency between the shear test (2D) and the falling ball test (3D) is achieved with good precision, while visible discrepancies may be due to special bondary conditions applied in the falling ball test. As a result, the numerical viscosity evaluated in shear test may be used as a good reference for other problems to apply adjustements.

\subsection{Simulation of the flow around cylinders}

\begin{figure}[t]
\centering
	\includegraphics[width=0.99\linewidth]{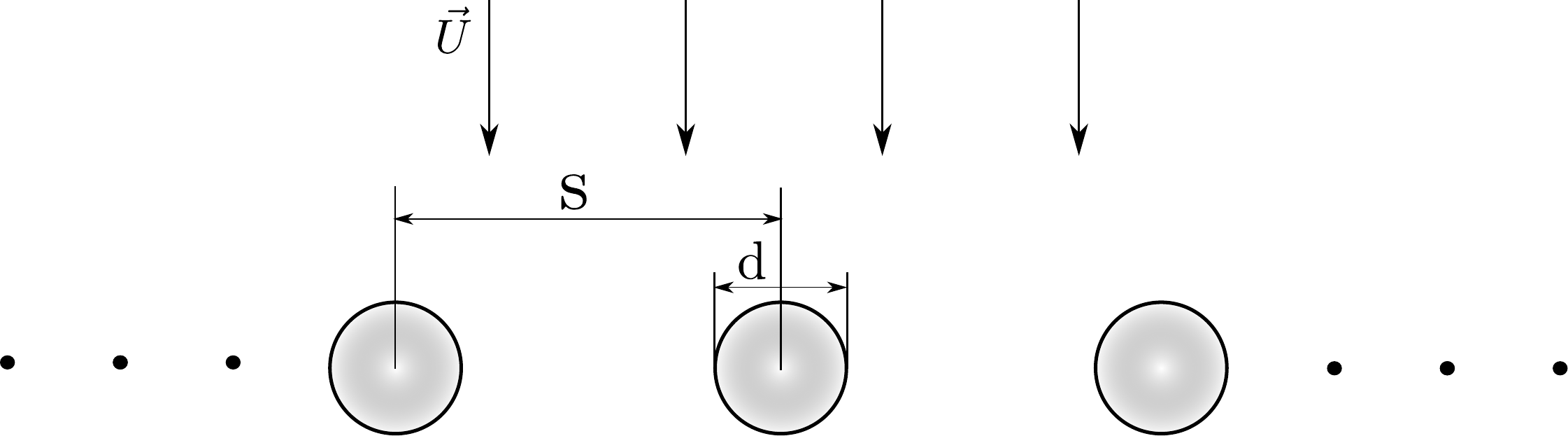}
	\caption{\label{cylinders-setup} The scheme of the flow around a set of cylinders: $d$ is the diameter of a cylinder, $S$ is the period. In our simulations a single cylinder in periodic boundary conditions in the appropriate domain is considered.}
\end{figure}

The flow of liquid at velocity $U$ passing through the cylinders of diameter $d$ placed with period $S$ between their centers pushes them with the drag force, as shown in Fig.~\ref{cylinders-setup}. The corresponding drag coefficient is defined as
\begin{equation*}
    C_D = \frac{F_D}{\frac{1}{2}\rho U^2d},
\end{equation*}
where $F_D$ is the drag force acting per unit length of the cylinder, $\rho$ is the density of the fluid.
According to \cite{ThiloMuller2014} the drag coefficient as a function of Reynolds number $Re$ can be determined by the formula
\begin{equation}\label{C_D}
    C_{D,M} = \frac{16\pi a_0}{Re},
\end{equation}
where
\begin{equation*}
    a_0 = \left[ 1 - 2\ln(2\omega) + \frac{2}{3}\omega^2 - \frac{1}{9}\omega^4 + \frac{8}{135}\omega^6  -   \frac{53}{1350}\omega^8 + \frac{1112}{42525}\omega^{10} -   \frac{241643}{13395375}\omega^{12} + \frac{18776}{1488375}\omega^{14} \right]^{-1},
\end{equation*}
\begin{equation*}
    \omega = \frac{\pi d}{2S}.
\end{equation*}

The proposed viscosity adjustment algorithm is applied to calculate the drag coefficients in SPH simulations for the paraffin and glycerol flows. In total, 6 simulations are performed: 3 without viscosity correction and 3 with it. Table~\ref{TableC_D} shows the obtained drag coefficients compared to the equation (\ref{C_D}). The Table~\ref{TableC_D} shows that the coefficients obtained with the viscous solver proposed in this paper are in good agreement with theoretical values~(\ref{C_D}).

\begin{figure}[t]
\centering
	\includegraphics[width=0.99\textwidth]{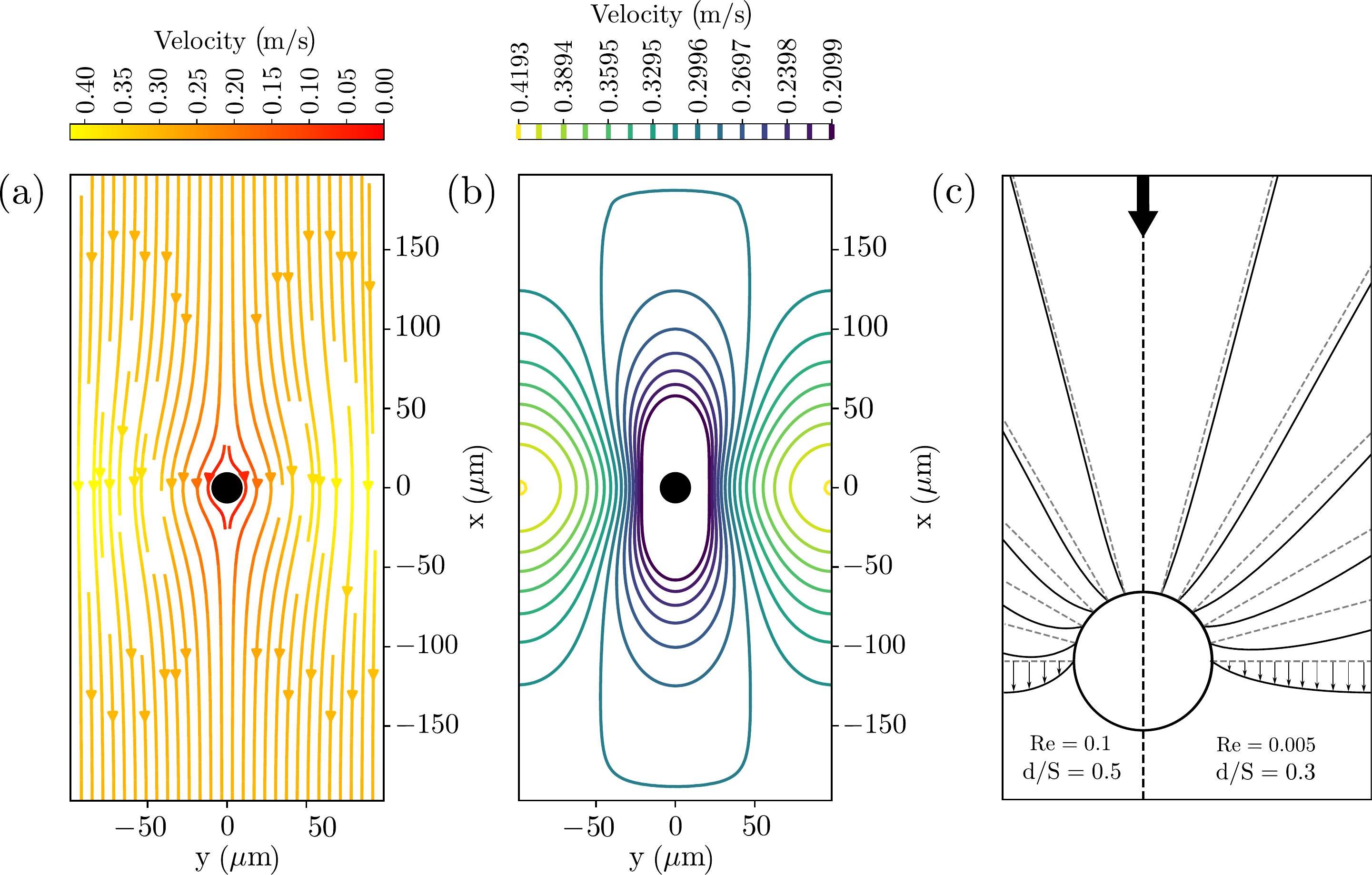}
	\caption{\label{velocity-field-cylinders} (a), (b) Velocity field around the cylinder at $d = 19.7333\,\mu m$, $d/S = 0.1$, $Re = 0.005$, $U = 0.3\,m/s$. (c) Normalized profiles of the tangential velocity component along different directions (gray dashed lines) at different Reynolds numbers and lattice parameters $S/d$. The tangential velocity component vectors originate perpendicularly from the shaded gray lines originating at angles 90$^o$, 75$^o$, 60$^o$, 45$^o$, 30$^o$, 15$^o$ to the positive direction of the $x$ axis. The flow direction is indicated by the arrow.}
\end{figure}

Figsures~\ref{velocity-field-cylinders}(a,b) show the velocity streamlines (a) and isolines (b) which are formed in liquid around a cylinder in periodic boundary conditions. Typical tangential velocity profiles for different Reynolds number values $Re$ are given in Fig.~\ref{velocity-field-cylinders}(c) which agree with ones obtained in~ \cite{ThiloMuller2014}.

\begin{table}
\caption{\label{TableC_D} Drag coefficients $C_D^{\mathrm{SPH}}$ obtained in SPH simulations with and without numerical viscosity correction compared to ones calculated using the equation~(\ref{C_D}).}
\centering
\renewcommand\arraystretch{1.3}
\def\tabrowsep{\noalign{\vskip 2pt}}
\begin{tabular}{ |c|c|c|c| }

\hline
$d/S$ &$C_{D,M}= \frac{16\pi a_0}{Re}$  & $\begin{smallmatrix} C_D^{\mathrm{SPH}}\\ \text{(corrected)}\end{smallmatrix}$ & $\begin{smallmatrix} C_D^{\mathrm{SPH}}\\ \text{(without correction)}\end{smallmatrix}$\\ \hline
\multicolumn{4}{|c|}{Paraffin, $Re  = 0.005$, $d = 0.1\,$mm, $U = 4.40\,$mm/s}\\\hline
$0.5$ & 20200   &  22800               & 50600                \\\hline
\multicolumn{4}{|c|}{Glycerol, $Re  = 0.005$, $d = 1\,$mm, $U = 5.92\,$mm/s}\\\hline
$0.5$ &20200   &  26900               & 68700                \\ \hline
$0.3$ &8000    & 10500                & 25900                \\ \hline
\end{tabular}
\end{table}

\begin{figure}[t]
\centering
	\includegraphics[width=1.0\textwidth]{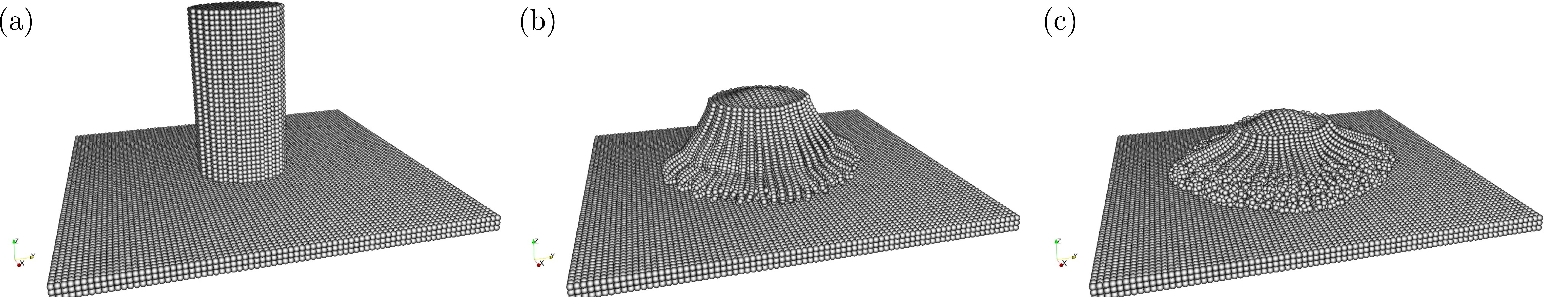}
	\caption {\label{liquid-cylinder-spreading} (a) The initial position of the cylindrical water drop with SPH particle size $D = 10^{-8}\,$m at $t=0\,$s. (b) Evolution of the drop without the numerical viscosity correction ($\zeta=0$). (c) Evolution of the drop with the numerical viscosity correction ($\zeta=1)$.
}
\end{figure}

\subsection{Spreading of a liquid cylinder on a rigid surface}

The qualitative test of the developed method of numerical viscosity control is the spreading of a cylindrical water drop on a rigid surface under the gravity force (without surface tension effects). The initial state of a liquid is shown in Fig.~\ref{liquid-cylinder-spreading}(a). Simulation without numerical viscosity correction ($\zeta=0$) is shown in Fig.~\ref{liquid-cylinder-spreading}(b), while Fig.~\ref{liquid-cylinder-spreading}(c) shows the simulation result for $\zeta=1$ at the same moment of time. One may clearly see that less viscous droplet spreads faster, which may be essential for realistic water simulations.

\section{Conclusion}

In this study we completed the formulation of contact SPH method by introducing the viscosity tensor to the governing equations. However, simulations of real fluids with the proposed method are hampered by the overwhelming numerical viscosity. The latter was found to be of similar nature as real viscosity does, which follows from the Riemann problem solution and is verified by the shear flow test simulations. As a result, the numerical viscosity may be subjected to a correction using similar terms as for the real viscosity in equations, thus providing the possibility of real fluids modeling.

By performing a set of shear flow simulations we demonstrated that the numerical viscosity linearly depends on the particle size and acoustic impedance of material. It was found, that numerical viscosity is comparable with the real one at particle sizes of $D \sim 10\,$nm, which are very small for real applications. Thus, the proposed viscosity adjustement algorithm becomes an essential tool in practice.

Despite of successful applications for single- and multi-material 2D shear flow, there is still doubt of the validity of the viscosity adjustement algorithm in 3D flows with complex geometry. To demonstrate the consistency of the developed method, we performed the falling ball test simulation, where the viscosity is determined using the Stokes method. It is demonstrated, that numerical viscosity obtained from the falling ball test is consistent with the one obtained in shear tests, so that the approximation~\eqref{eq:numerical-viscosity-approximation} may be used to estimate the numerical viscosity for various applications.

It was shown that the application of the numerical viscosity correction allows to predict drag coefficients for the flow around cylinders in accordance with theoretical predictions. The test with liquid drop spreading on the surface qualitatively demostrate the effect of numerical viscosity reduction in real-life simulations.

The proposed contact SPH method is completed for the evaluation of compressible viscous, elastic-plastic, heat conductive media.

\bibliography{}

\end{document}